\documentstyle[12pt,epsf,epsfig,amsmath,amsfonts,color]{article}

\def\be{\begin{equation}}
\def\ee{\end{equation}}
\def\ba{\begin{eqnarray}}
\def\ea{\end{eqnarray}}

\def\bdm{\begin{displaymath}}
\def\edm{\end{displaymath}}
\def\la{~\mbox{\raisebox{-.6ex}{$\stackrel{<}{\sim}$}}~}
\def\ga{~\mbox{\raisebox{-.6ex}{$\stackrel{>}{\sim}$}}~}
\def\bq{\begin{quote}}
\def\eq{\end{quote}}

 at 10truept

\newcommand{\beq}{\begin{equation}}
\newcommand{\eeq}{\end{equation}}
\newcommand{\bea}{\begin{eqnarray}}
\newcommand{\eea}{\end{eqnarray}}
\newcommand{\beqa}{\begin{eqnarray}}
\newcommand{\eeqa}{\end{eqnarray}}

\newcommand{\mpl}{{M_{\rm Pl}}}

\def\la{~\mbox{\raisebox{-.6ex}{$\stackrel{<}{\sim}$}}~}
\def\ga{~\mbox{\raisebox{-.6ex}{$\stackrel{>}{\sim}$}}~}

\def\ltap{\ \raise.3ex\hbox{$<$\kern-.75em\lower1ex\hbox{$\sim$}}\ }
\def\gtap{\ \raise.3ex\hbox{$>$\kern-.75em\lower1ex\hbox{$\sim$}}\ }
\def\gl{\ \raise.5ex\hbox{$>$}\kern-.8em\lower.5ex\hbox{$<$}\ }
\def\roughly#1{\raise.3ex\hbox{$#1$\kern-.75em\lower1ex\hbox{$\sim$}}}

\begin{document}

\thispagestyle{empty}
\begin{flushright}
April 2024 \\
DESY-24--040
\end{flushright}
\vspace*{1.25cm}
\begin{center}

{\Large \bf Falsifying Anthropics}

\vspace*{1cm} {\large 
Nemanja Kaloper$^{a, }$\footnote{\tt
kaloper@physics.ucdavis.edu} and 
Alexander Westphal$^{b,}$\footnote{\tt alexander.westphal@desy.de}}\\
\vspace{.2cm} 
{\em $^a$QMAP, Department of Physics and Astronomy, University of
California}\\
\vspace{.05cm}{\em Davis, CA 95616, USA}\\
\vspace{.2cm} $^b${\em Deutsches Elektronen-Synchrotron DESY, Notkestr. 85, 22607 Hamburg, Germany}\\

\vspace{1.5cm} ABSTRACT
\end{center}
We propose using fuzzy axion dark matter to test the anthropic principle. 
A very light axion can be directly detectable, at least by black hole superradiance effects. 
The idea then is that gravitational and astrophysical observations can 
discover a light axion in the regime where it must be all of dark matter with abundance
which must be set up by the anthropic principle, due to excessive primordial misalignment induced by 
inflation-induced Brownian drift of fluctuations. 
Yet it may turn out that dark matter is something else instead of this axion. Since the de Sitter-induced 
axion misalignment controlled only by the de Sitter curvature cannot be evaded, 
this would invalidate the anthropic prediction of the dark matter abundance.

\vfill \setcounter{page}{0} \setcounter{footnote}{0}

\vspace{1cm}
\newpage

\hfill \parbox{12.37cm}{
{\it  A physicist talking about the anthropic principle runs the same risk}\\
{\it as a cleric talking about pornography: no matter how much you say}\\
{\it  you're against it, some people will think you're a little too interested...} }
\begin{flushright}
\hfil{{\it (attributed to S. Weinberg)} \cite{dine}}
\end{flushright}

\vskip0.5cm

The anthropic principle \cite{dicke,carter,barrowtipler} is often invoked as a ``last ditch" tactic to account why certain quantities 
take the values they do, when those seem unnatural or mysterious from the microscopic and/or dynamical point of view.
Frequently encountered in cosmology, in applications to cosmic origins \cite{tryon,hawkingcollins}, cosmological parameter
determination such as cosmological constant \cite{weinberg} and dark matter (DM) abundance \cite{linde}, estimates of extraterrestrial
life \cite{drake}, and more recently even standard model parameters \cite{savas,giudice,Kaloper:2019xfj}, sometimes anthropic reasoning 
is set aside when new ideas for dynamics arise. For example, the anthropic explanations of observed size and isotropy of the
universe \cite{tryon,hawkingcollins} were reneged upon after the advent of cosmic inflation. In other examples, 
such as the determination of the cosmological constant \cite{weinberg}, 
anthropics is still viewed as a viable explanation. 

On the other hand, it can 
happen that Nature intercedes against a dynamical explanation of the observed value of a physical quantity, 
such as in the example of WIMP DM, which may not exist despite the fact that if it did it would 
naturally explain the required DM abundance. If so, this can open 
the stage to the anthropic reasoning, which was not evoked previously and 
which might be required to explain the DM abundance if the DM 
is an ultralight field, such as e.g. an axion \cite{linde,linde2}. These examples illustrate an outstanding 
dilemma about the anthropic reasoning, that concerns its predictivity, ranging between the extreme applicability and complete
irrelevance, as discussed in, e.g. \cite{lenny,smolin}. 

An obvious, and simple-sounding question concerns establishing the positivistic criteria for the applicability of the anthropic reasoning. 
This issue has been very elusive, which has led some to even proclaim anthropics to be outside of the domain of science, alleging that it
is a tautology (see, e.g. \cite{smolin}). The existing considerations, to the best of our knowledge, do not provide a clear cut case 
for how the anthropic reasoning can be falsified. Specifically, a setup where anthropic predictions are directly excluded by an 
observation has not been available to date.  It is our purpose to provide such an example in this work. 

In a nutshell, our argument proceeds as follows. Let us imagine that our universe emerges from an 
inflationary multiverse, which includes an era of rapid early inflation
at some high scale $V^{1/4} \simeq \sqrt{\mpl H_I}$. In the simplest realization of such a scenario, 
where all the required $\sim 60$ efolds of inflation
occur in one go, and the density perturbations which seed the late stage structure formation (galaxies, clusters etc) originate as
quantum zero point fluctuations of the inflaton field, the scale of inflation should be very high, $H_I \la 10^{13} \, {\rm GeV}$, or 
$V^{1/4} \la M_{\rm GUT} \sim 10^{16} \, {\rm GeV}$ \cite{Linde:1990flp}. 
This scenario can be probed, and eventually (still!) confirmed, by observations. For example,
the LiteBIRD satellite \cite{LiteBIRD:2022cnt} could confirm this by discovering the imprint of the primordial gravity waves
in the cosmic microwave background, with the tensor to scalar power
ratio $r \simeq 0.001$, which is correlated with the scalar perturbations, 
and in particular with the scalar spectral index $n_{\rm S} \simeq 0.965$. The specific models which
could explain this are not too important here, but we can mention either the Starobinsky inflation and its variants 
\cite{Starobinsky:1980te}, or variants of monodromy models \cite{awes,kalsor} 
with short interruptions \cite{DAmico:2020euu,DAmico:2021vka,DAmico:2021fhz}. Both are still viable. 
Such scenarios can be accommodated naturally in the landscape frameworks with many very light fields, which 
arise due to approximate shift symmetries, that come from 
mixing with fluxes in string compactifications. 

In this case an axion would be a natural ultralight DM candidate (see, e.g. 
\cite{Hu:2000ke,Cicoli:2012sz,Hui:2016ltb}). In the minimal framework, its presence could be confirmed by black hole superradiance, 
which is a universal phenomenon thanks to the universality of gravity and the Equivalence Principle. 
Observations of supermassive black hole superradiance \cite{Arvanitaki:2009fg,Arvanitaki:2010sy} could then determine the
axion mass $m_\varphi$ and decay constant $f_\varphi$. The 
former could in principle span a broad range of values, while the latter is expected to be
$\mpl \ga f_\varphi \ga M_{\rm GUT}$ on theoretical grounds (see, e.g. \cite{Svrcek:2006yi}). 
In this regime of scales, the DM axion would automatically have the correct relic abundance if the mass is 
$m_\varphi \sim 10^{-19} \, {\rm eV}$ for $f_\varphi \sim M_{\rm GUT} \sim 10^{16} \, {\rm GeV}$, 
which is very close to the current bound $m_\varphi \ga {\rm few} \times 10^{-20} \, {\rm eV}$ 
\cite{Rogers:2020ltq}. This ``goldilocks" value could then be excluded by future superradiance searches: these could discover 
the ultralight axions with masses $m_\varphi > 10^{-19} \, {\rm eV}$ and $f_\varphi \ga M_{\rm GUT}$. 

By universality of gravity and universality of inflation, an existence of a DM axion 
in this range would make it {\it unavoidable} that this axion in fact {\it must} be all of DM, with abundance set by anthropics 
to avoid overclosing the universe. The overclosure problem has anthropic implications, because too much axion
would yield too much structure formation or force a change in the baryon abundance, affecting late cosmology 
in ways adversarial to emergence of life \cite{linde,linde2,wilczek,Hellerman:2005yi,Freivogel:2008qc,nkter}.
In the minimal framework of local physics, there just is no other way to avoid these problems but to deploy
anthropic principle, as we will see below. 
This would imply that we live in a special corner of the multiverse where axion DM
tends to naturally alter structure formation, and where we must pick the special corner where structure forms by
anthropic reasoning \cite{linde,linde2,wilczek,Hellerman:2005yi,Freivogel:2008qc,nkter}. 

And then, \underbar{direct DM observations could reveal that DM is {\it not} an axion}, and not even an ultra-light field! 
In the minimal framework, this would contradict the anthropic arguments, because the axion abundance 
would have to be much smaller than the anthropically determined value, meaning that we live in a very special
corner of the multiverse, which is very unlikely in the anthropic sense. Some other dynamics had to 
intervene to select it -- and the anthropic principle simply did not deliver. 

\begin{figure}[thb]
 \centering
 \hspace{2cm}\includegraphics[width = 0.4 \textwidth]{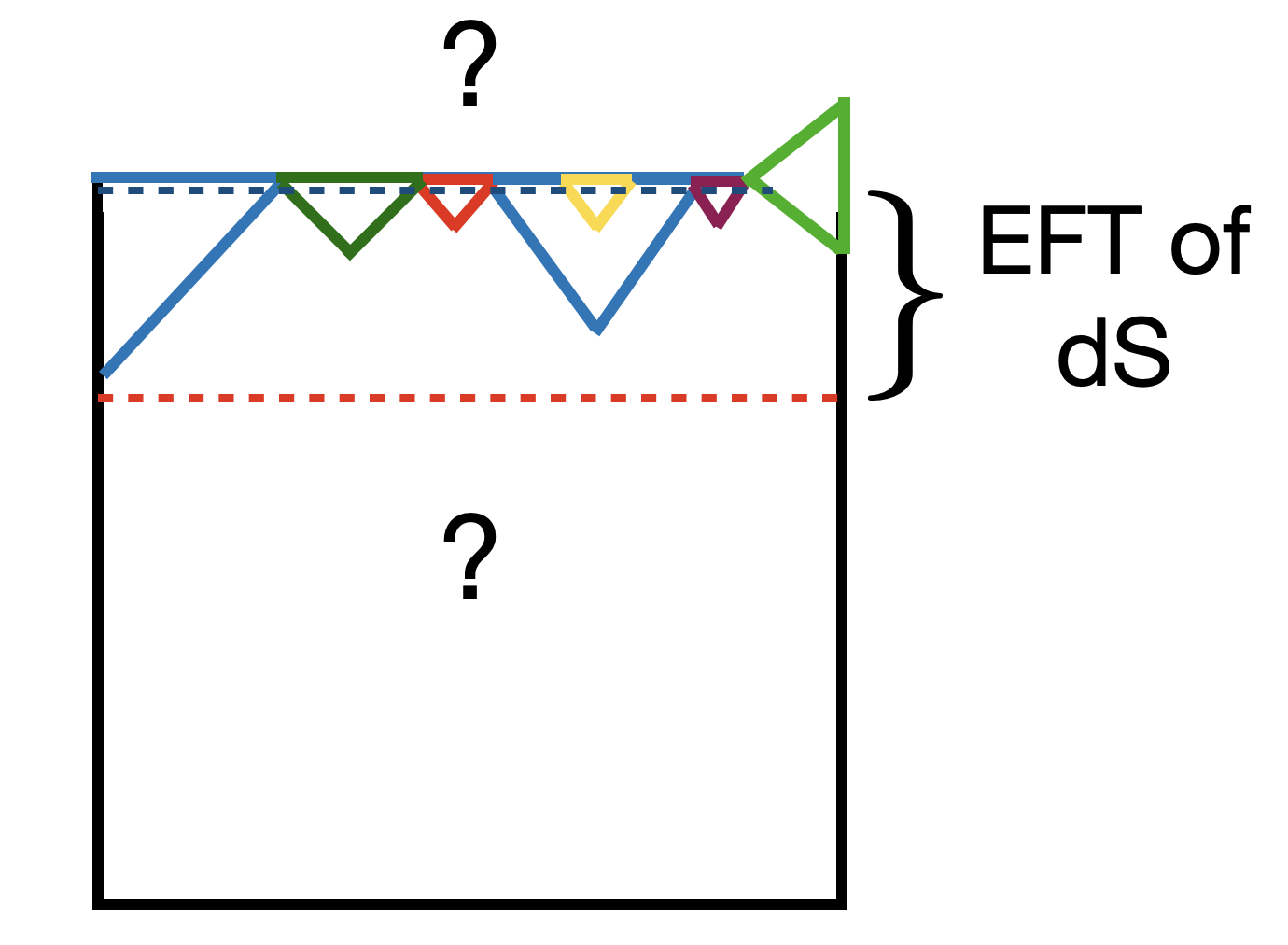}
 \caption{dS ``bubble-verse".}
  \label{fig:dSverse}
\end{figure}
Let us now flesh out the details. 
We suppose the universe was created by inflation: at some high scale the 
universe started as a patch of a (quasi)dS ``bubble-verse", whose Penrose diagram ``cartoon" is provided in
Fig. (\ref{fig:dSverse}). This picture is emerges naturally from the cosmological constant screening mechanisms which
employ the space-time filling fluxes of $4$-forms, with stress energy which is locally degenerate with the cosmological 
constant, but can change by discharges mediated by 
Schwinger processes \cite{brown,bp,Kaloper:2022yiw,Kaloper:2022jpv,Kaloper:2023kua}. Regardless
of the criteria for the selection of the terminal value of the cosmological constant, the background space-time looks like 
a fractalized dS due to the evolution of bubbles bounded by charged membranes, which alter the local value
of the cosmological constant. In some regions of this ``bubble-verse" the evolution will find a way to an inflationary trajectory,
which leads to reheating and late time structure formation, that matches our corner of the multiverse. 
The complete theory of a bubbly multiverse is not yet available. Yet, we have a good description of the dynamics 
in the region denoted by the dashed lines in Fig. (\ref{fig:dSverse}), where bubble
creation and evolution in the background dS occurs, based on standard GR, locality and causality. This is 
the semiclassical regime of QFT coupled to GR. The unknowns in this
picture are mostly related to the issues of initial and final conditions, depicted by the question marks in Fig. (\ref{fig:dSverse}). 
The dynamics which our arguments here rely on are generically independent of those, and completely fall in the regime between
the dashed lines. Thus we will ignore the unknowns in Fig. (\ref{fig:dSverse}) and proceed by analyzing the evolution in the
semiclassical regime. 

Next we suppose that there exists an ultra-light scalar field, which will eventually play the role of fuzzy DM.
This field is very light during inflation, and so it quickly ``freezes in" due to the cosmological expansion, since
it will be in slow roll. Thus in a typical bubble with a large local cosmological constant it will be merely a spectator early on.
Its dynamics is controlled by the action
\be
S \supset \int d^4x\sqrt{g}\left(\frac12 (\partial\varphi)^2-\frac{m_\varphi^2}{2}\varphi^2\right) \, . 
\label{scalaract}
\ee
Here, quantitatively ``ultra-light" means that $m_\varphi \ll H_{I}$, where as above, $H_I$ is the Hubble scale during inflation. 

\begin{figure}[thb]
 \centering
 \includegraphics[width = 0.57\textwidth]{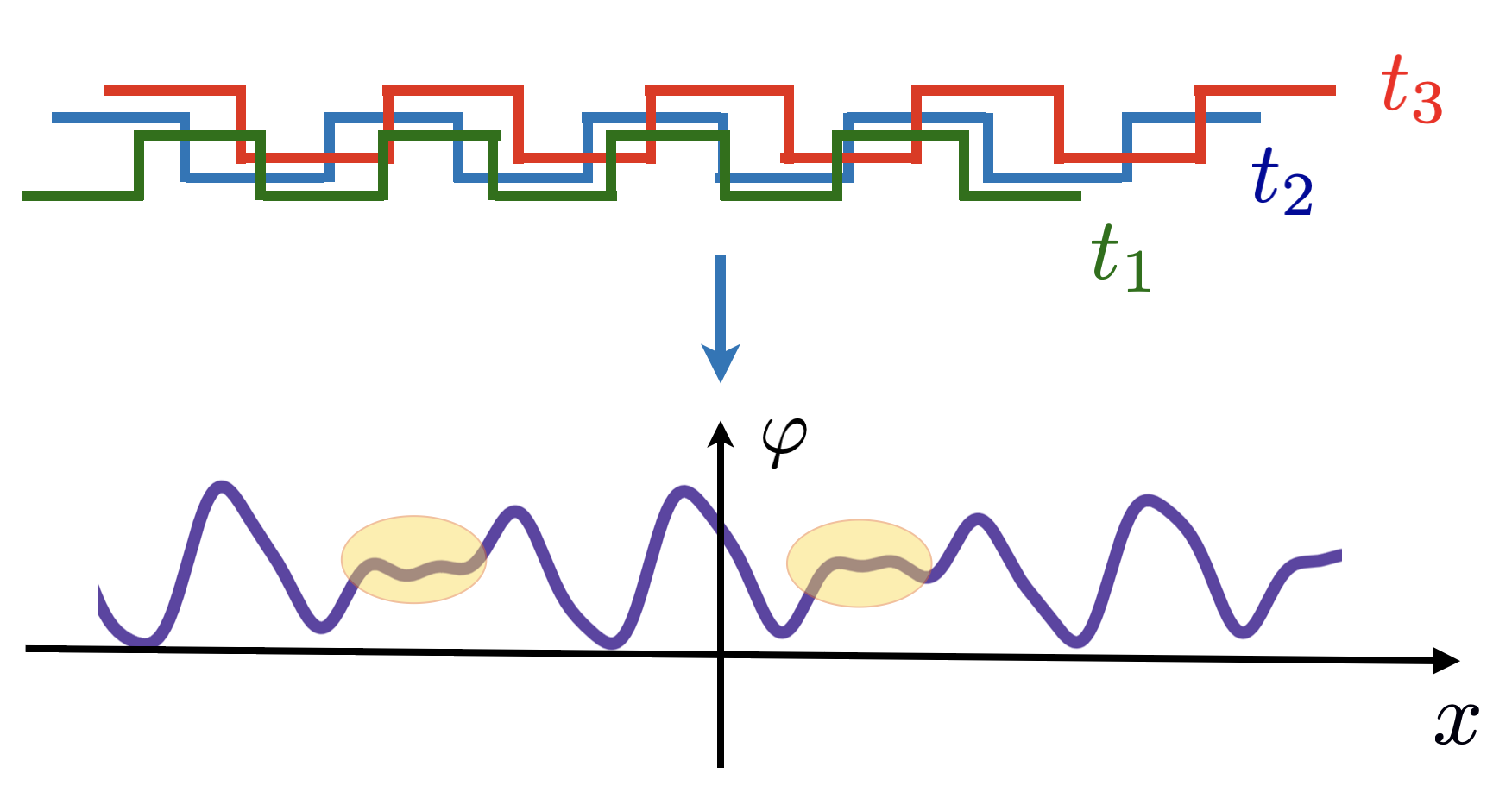}
 \caption{dS random walk.}
  \label{fig:dSjumps}
\end{figure}

Classically, such an ultra-light field would get arrested by background expansion rapidly. Within a few Hubble times
$\sim H^{-1}_I$, it would become a constant. However, quantum-mechanically, the ``constant" would be anything but: due to the
quantum fluctuations, the field background value would get corrected by fluctuations at all scales, which would
approach a constant once their wavelength becomes larger than the horizon size $\sim H_I^{-1}$ \cite{Linde:1990flp,Mukhanov:1981xt}.
These processes are illustrated in Fig. (\ref{fig:dSjumps}),
and involve the generation of fluctuations of $\varphi$ due to Heisenberg's uncertainty principle, which get stretched by the background 
quasi-dS geometry, and freeze once the wavelength exceeds $H^{-1}_I$. The process continues as long as the background remains
close to dS, and in any fixed location in dS, the generation and subsequent freeze-out will correct the background geometry due to
the superposition of the fluctuations with random relative phases. 

The actual process of evolution of these fluctuations involves the standard de Sitter space zero-point fluctuations, which 
in momentum domain involve the usual fluctuations of the momentum space Fourier modes $\hat \varphi_{\vec k} \sim H_I$. These modes
initially have small wavelength which stretches by de Sitter expansion until it 
reaches the horizon $\sim H_I^{-1}$. At this point the momentum mode amplitude freezes at a constant value set by $H_I$. 
The process repeats over the full range of scales as long as the background remains quasi-de Sitter, and such modes 
are continually produced until the end of inflation. As they freeze out, 
from the point of view of subhorizon observers they appear as corrections on top of the homogeneous background, which 
add up with a random phase. Because of this randomness, 
somewhere the fluctuations will cancel out, and somewhere 
they will add up coherently, and behave as a correction to the background over the whole Hubble patch (depicted by golden-shaded 
regions in Fig. (\ref{fig:dSjumps})). The momentum domain modes adding up to the background will renormalize the 
coordinate domain {\it vev} $\varphi$, and make it grow during the inflationary stage. 
This is of course the famous IR instability of dS space \cite{Vilenkin:1983xp,Linde:1990flp,Tsamis:1994ca}. 
Due to the quantum fluctuations, the most likely value of the vacuum expectation of the 
ultra-light field will not be zero, $\langle\varphi\rangle \neq 0$. Instead,
the growth of the field will go on until the field energy density reaches
$\rho_\varphi \simeq \langle \dot \varphi^2 + (\vec \nabla \varphi)^2+m^2\varphi^2 \rangle_{dS} \sim H_I^4$, 
saturating the de Sitter space IR cutoff given by $H_I$. Note that this is merely an aspect of the fact 
that in dS the physical expectation values of quantum operators undergo
random evolution, varying from a patch to a patch, drifting around due to quantum effects. The patches where the coordinate 
domain field {\it vev} is such the energy density is $H_I^4$ are the most likely ones, as is well known from some classic works
\cite{Parker:1968mv,Bunch:1978yq,Vilenkin:1982wt,Linde:1982uu,
Vilenkin:1983xp,Linde:1990flp,Starobinsky:1994bd}, which studied this phenomenon in much more 
detail using renormalization of quantum field theory in curved spacetime. 

Therefore focusing on the typical case we can  estimate the scalar {\it vev} drift 
employing gravitational particle production rates. To estimate the typical value of the scalar {\it vev} 
we use the estimate of the typical value of energy density stored in the 
slowly rolling scalar field {\it vev} in coordinate domain, ignoring inhomogeneities and slow roll corrections, and ignoring the
precise values of numerical coefficients of ${\cal O}(1)$ \cite{Bunch:1978yq,Vilenkin:1983xp,Linde:1990flp},
\be
m_\varphi^2 \langle \varphi^2 \rangle \sim H_I^4 \, .
\label{energy}
\ee
This immediately yields 
\be
 \langle\varphi\rangle\sim \sqrt{\langle \varphi^2 \rangle} \sim \frac{H_I^2}{m_\varphi} \, .
\label{vev} 
\ee
Clearly, when $m_\varphi \ll H_I$, the {\it vev} $\langle \varphi \rangle$ will be huge. Using $V_I \simeq \mpl^2 H_I^2$  
\be
\frac{\langle\varphi\rangle}{\mpl} \sim \frac{H^2_I}{\mpl m_\varphi}  \simeq \frac{V_I}{M_{\rm P}^4} \frac{\mpl}{m_\varphi} \gg 1 \, ,
\label{vevhuge}
\ee
when the scale of inflation is high, close to $M_{\rm GUT}$, and $m_\varphi \ll H_I$. 
In fact, for single field inflation, with tensor to scalar ratio which 
saturates the current bound of $r \la 10^{-2}$, we can estimate that
$(H_I/\mpl)^2 \sim r \bigl(\frac{\delta \rho}{\rho}|_{\rm S}\bigr)^2 \sim 10^{-12}$. 
Thus, any scalar whose mass is smaller than
\be
m_* \simeq 10^{-12} \mpl \simeq 10^{15} \, {\rm eV} \, ,
\label{masscritical}
\ee
and which is defined over a real line, or a very large field space 
with super-Planckian perimeter or period, will invariably pick up a super-Planckian {\it vev}, 
$\langle \varphi \rangle > \mpl$, driven by quantum fluctuations. 

\begin{figure}[thb]
 \centering
 \includegraphics[width = 0.57 \textwidth]{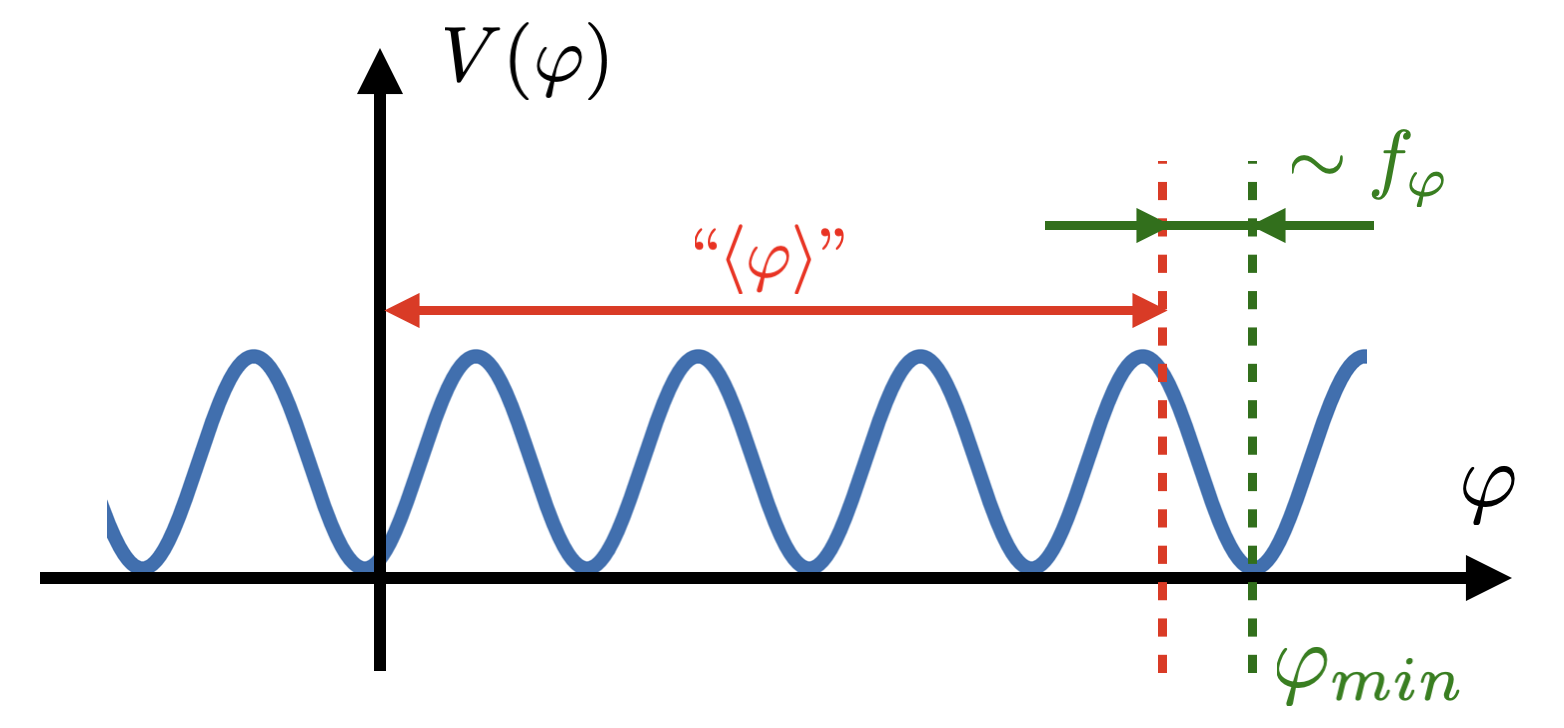}
 \caption{dS random walk for compact pseudo scalar field.}
  \label{fig:dSjumps2}
\end{figure}

This is too much, indicating a pathology with non-compact $\varphi$, 
according to the current ideations about quantum gravity \cite{Arkani-Hamed:2006emk}. A simple ``cure" to this problem is
to only use those field theories which 
involve only compact scalars with sub-Planckian periods $f_\varphi < M_{\rm P}$, when seeking a description 
of the multiverse from which our universe can spring up. In this case, the preceding argument shows 
that the typical field {\it vev} will be $\langle\varphi\rangle \sim f_\varphi$: the largest it can be. This is illustrated in 
Fig. (\ref{fig:dSjumps2}), showing that due to the periodicity of the field space, the super-Planckian field ranges are 
merely a mirage, and that the typical field misalignment from a vacuum is of the order of $f_\varphi$. 
Arguably, the simplest frameworks which realize such dynamics and give rise to compact scalars 
are variants of the axiverse \cite{Arvanitaki:2009fg}, where the compact fields are axions which 
get their masses from nonperturbative gauge dynamics in 
strong coupling regime, that are therefore naturally small. We shall work with this premise from here on. 
We will use the label ``scalar" for it, although technically we will mean ``pseudo-scalar". 

If dS (or a smooth inflationary background) were eternal, never ending, the fluctuations  
wouldn't matter so much. The {\it vev} of $\varphi$ would just be another landscape variable, forever frozen in an eternally structureless
and empty universe. But dS seems to end, e.g. via flux discharges that reduce 
the cosmological constant \cite{brown,bp,Kaloper:2022yiw,Kaloper:2022jpv,Kaloper:2023kua}. In fact, slow roll inflation may very well 
be the last stage of relaxing a large $\Lambda$ to the tiny terminal value. So $\langle\varphi\rangle$ condensate formed by the 
Brownian drift induced by background dS geometry will eventually melt and decay. These processes involve several important stages,
which we should consider carefully since our goal is to ultimate estimate the effect of the melted scalar ``permafrost" $\langle \varphi \rangle$
on the background geometry. 

First off, when inflation ends, since $\varphi$ is ultralight, $m_\varphi \ll H_I$, and subleading, with typical initial energy density 
$\rho_\varphi \simeq m^2_\varphi f^2_\varphi \ll H_I^2 \mpl^2$ (since $f_\varphi < \mpl$), this scalar spectator will remain in slow 
roll for a long time. It will behave as a contribution to residual (``early") dark energy, which however decays once $H < m_\varphi$. 
Thereafter, $\langle \varphi \rangle$ will slosh around the minimum $\varphi=0$ (which is chosen to be at zero for convenience, by 
shifting the scalar by an integer number of periods). The scalar energy density from this moment on will behave as 
DM, $\rho_\varphi\sim a^{-3}$. To summarize,
\be
\rho_\varphi \simeq \begin{cases}
m_\varphi^2 f_\varphi^2 \, , 
~~~~~~~~~~~~~  {\rm for~~} H > m_\varphi :~~{\rm ``early" ~dark ~energy} \, ;\\
\\
m_\varphi^2f_\varphi^2 \left(\frac{a_0}{a}\right)^3 \, ,  ~
~~~~  {\rm for~~} H > m_\varphi :~~{\rm ``fuzzy" ~dark ~matter}\, .
 \end{cases}
\label{ededm}
\ee
We refer to the dark matter as ``fuzzy" since the field is ultra-light, $m_\varphi \ll H_I$. 
To actually get the precise normalization of the scale of fuzzy DM abundance, we need to determine the epoch of 
EDE to fuzzy DM transition, i.e. the permafrost meltdown era. This is encoded by the parameter $a_0$ in Eq. (\ref{ededm}). 
In general, to do this we would need to specify the detailed thermal history of the universe from today to the end of inflation. However
such precision is not required for our argument; it suffices to work within an order of magnitude. Hence 
for simplicity we can assume that inflation was followed by radiation epoch right away.

Note that the abundance of this fuzzy dark matter, for a fixed axion decay constant $f_\varphi$, depends on the square
of the mass of the axion. This means that the axion DM contents is quite sensitive to the mass. If a mass were 
chosen to precisely reproduce all of the DM in the universe today, changing it by only a factor of a few would 
make the axion either irrelevant as a DM candidate,
or it would imply that it gives way too much DM. This sensitivity is the central 
reason for our argument, as we will see below.

So, right after inflation, we take the universe to be  radiation-dominated, with energy density $\rho \simeq T_{reh}^4 /{a}^4$,
where we take the scale factor at reheating to be unity, $a_{\rm reheating} = 1$, employing residual lapse function reparametrization gauge 
symmetry. After that, for as long as $m_\varphi < H$, the scalar remains frozen. Once $H= m_\varphi$ the scalar permafrost melts. At that
instant,
\be
m_\varphi^2M_{\rm P}^2 \simeq T_{reh}^4 \left(\frac{1}{a_{0}}\right)^4\;\;\Rightarrow\;\; a_{0} \simeq \frac{T_{reh}}{\sqrt{m_\varphi M_{\rm P}}}
\label{meltdown}
\ee
From this moment onward, the fuzzy DM, which is initially subdominant, begins to chase after the dominant radiation since 
$\rho_\varphi \simeq m_\varphi^2f_\varphi^2 (a_{0} /a)^3$ and $\rho_{rad} \simeq (T_{reh}/a)^4$, and so 
the radiation dilutes faster during expansion. Eventually, and invariably, DM catches up to 
radiation, at the temperature $T_\star$ determined by $\rho_{rad}=\rho_\varphi$
\be
T_\star=\frac{T_{reh}}{a} \simeq \frac{m_\varphi^{1/2} f_\varphi^2}{\mpl^{3/2}} \, .
\label{tstar}
\ee
A variant of this equation was given previously in \cite{Cicoli:2012sz}. 
In principle, the value of $T_\star$ is an arbitrary multiverse parameter, which is controlled by the initial conditions. However, in our universe
it is bounded -- by observations. We know that in our universe, the nonrelativistic matter was subdominant to radiation until 
radiation-matter equality, which occurred when the CMB temperature was about $T_{rm} \simeq {\rm eV}$. As a result, there are implications 
for the massive scalar dynamics, which range from irrelevant to extremely constraining, depending on the values of $m_\varphi, f_\varphi$. 
The following options can occur:
\begin{itemize}
\item the ultra-light scalar is a component of DM, but its mass and/or decay constant are too small, yielding $T_\star \ll {\rm eV}$, and so 
it is subdominant, thus being mostly irrelevant;
\item this DM component is an ${\cal O}(1)$ fraction of our DM by luck; i.e this axion is 
a ``goldilocks" field, with just the right mass and decay constant to give $T_\star \sim {\rm eV}$, and so to 
be all of currently observed DM in the universe; 
\item this DM component may be our DM, but its mass $m_\varphi$  is too large; with a given $f_\varphi \simeq 
M_{\rm GUT}$, this DM would give $T_\star \gg {\rm eV}$, 
and so it would overclose the universe if its misalignment from the vacuum is given by the
typical, natural value in dS\footnote{It was suggested in \cite{Cicoli:2021gss} that this might be viewed as a constraint on
landscape model building.}; therefore, its observed misalignment would have to be smaller, and hence atypical, 
that without fine tunings could only be understood anthropically in the minimal multiverse 
framework which we deploy here; this DM would then \underbar{have to be all of the observed
DM} for the anthropic argument to work. 
\end{itemize}

To get a quantitative idea on these three options for fuzzy DM, 
recall that string axiverse model building appears to favor $f_\varphi\sim M_{\rm GUT}$ \cite{Svrcek:2006yi}. Then
the ``goldilocks" DM for which $T_\star \sim {\rm eV}$ corresponds to $m_\varphi\sim 10^{-19}\, {\rm eV}$, which is very close to
the bound on fuzzy DM mass $m_\varphi \ga {\rm few} \times 10^{-20} \, {\rm eV}$ \cite{Rogers:2020ltq}. This leads to 
what we consider an extremely curious possibility, of having future observations falsifying the anthropic reasoning. 
The logic is as follows:
\begin{enumerate}
\item next generation of CMB observations, specifically LiteBIRD \cite{LiteBIRD:2022cnt}, could 
essentially confirm inflation -- 
a period of quasi-dS expansion at a high scale -- by an observation of the tensor to scalar power ratio $r\simeq 0.001$ correlated 
with the scalar spectral index, with sufficiently precise determination of both quantities; while one might worry about whether 
minimal assumptions on which such a scenario is based are unique, these observations would pretty much nail the  
dynamics for most practical intents and purposes;
\item black hole superradiance observations could confirm the presence of an ultra-light axion in the universe, because 
superradiance and gravity are universal phenomena; the ``sweet spot" for sensitive superradiance observations 
is given by the equality between the axion mass and black hole's Hawking temperature,
$m_\varphi\sim T_{BH}\sim {\mpl^2}/{M_{BH}}$, which means that for the ``goldilocks" value of the axion mass,
$m_\varphi \sim 10^{-19}\, {\rm eV}$, the black hole mass should be $M_{BH} \sim 10^7 M_{\odot}$; this means, that
observing superradiance for black holes of mass below $10^7$ solar masses, the fuzzy axion would be heavy enough to 
overclose the universe if $f_\varphi \sim M_{\rm GUT}$; the latter parameter can also be checked by superradiance, by considering
superradiant cloud instabilities \cite{Arvanitaki:2009fg,Arvanitaki:2010sy}; 
\item by universality of inflation and existence of an ultra-light fuzzy DM axion 
in this range, with all the usual premises of minimal semiclassical gravity, this axion would have to be all of DM as we discussed
above\footnote{In this discussion we have ignored the possibility that inflationary generation of axion misalignment could also
yield a large amplitude of isocurvature perturbations which are constrained by CMB observations \cite{Linde:1985yf,Fox:2004kb,Hertzberg:2008wr,Mack:2009hs,Choi:2014uaa,Choi:2015zra}, because these are not in conflict with
anthropic reasoning, our primary interest here. The point of principle, that anthropics can be observationally 
falsified therefore remains unaffected. An attentive reader might complain that the current observational bounds on
isocurvature perturbations might be in conflict with falsifying anthropics in {\it our} universe; however, slightly nonminimal frameworks 
\cite{Choi:2014uaa,Choi:2015zra} illustrate this could be circumvented by matter sector dynamics.}; 
in particular, to avoid overclosing the universe, the axion abundance \underbar{must} be set anthropically; 
in other words, in this case we ended up living in a special corner of the multiverse where fuzzy axion DM
tends to naturally alter structure formation;
\item and then, finally, direct DM observations reveal that DM is \underbar{not} an axion and 
not even an ultra-light field; this means, a different dynamics produces DM and sets the right 
relic abundance, while the ultra-light axion is completely blocked from acquiring its natural 
anthropic value by pure chance, unlikely as it is: the axion is 
far more suppressed; this would imply, that our universe is a very atypical, very 
special corner of the multiverse, which would give a serious challenge to anthropic reasoning.
\end{enumerate}

To us, what is the most interesting here is that this is an in-principle experimental falsification of anthropic predictions, 
despite many past claims that anthropics cannot be falsified. It is theoretically conceivable that the framework 
needed to describe the universe
could in fact be very far from minimal, by which we take the combination of local 
QFT coupled to universal gravity, with de Sitter-like inflation and Brownian drift dynamics 
induced by geometry, and standard quantum mechanics governing local microscopic degrees of freedom. 
It could be that these premises are wrong in the real world, since either inflation, QFT, dS, GR, or something else is wrong.  
At least to us, it seems that hiding the tell-tale signs of nonminimal distortions of these premises from other observational 
tests would be extremely challenging. 
All of these caveats are higher order concerns in our view, and we do not think that we need to go there yet. 

But even more importantly, we think that even if we retain these assumptions, the fact that there exists a situation based on 
them that leads to a possibility of falsifying anthropics by making a prediction that does not come true is telling enough. 
So, in sum, we infer that the sequence of observations we outlined here would amount to effectively falsifying anthropics, 
proving that the anthropic principle is not a tautology. 

\vskip.5cm

{\bf Acknowledgments}: 
We thank G. D'Amico and J. Leedom for useful discussions. 
NK is supported in part by the DOE Grant DE-SC0009999. 
AW is partially supported by the Deutsche Forschungsgemeinschaft under Germany's 
Excellence Strategy - EXC 2121 ``Quantum Universe" - 390833306. This 
research was supported in part by grant NSF PHY-2309135 to the Kavli Institute for Theoretical Physics (KITP).

\end{document}